\documentclass[showpacs,prl,letter,fleqn]{revtex4}
\usepackage{amsmath,amssymb,amsthm}
\usepackage{epsfig}
\usepackage{graphicx}

\begin{document}

\title{Higher Order Stability of a Radiatively Induced $220$ GeV Higgs Mass}
\author{V. Elias$^{a,b}$, R. B. Mann$^{b,c}$, D. G. C. McKeon$^{a,b}$ and T.
G. Steele$^{d}$}
\affiliation{$^{a}$Department of Applied Mathematics, The University of Western Ontario,
London, Ontario N6A 5B7, Canada \\
$^{b}$Perimeter Institute, 31 Caroline Street North Waterloo, Ontario Canada
N2L 2Y5\\
$^{c}$Department of Physics, University of Waterloo, Waterloo, Ontario
Canada N2L 3G1\\
$^{d}$Department of Physics and Engineering Physics, University of
Saskatchewan, Saskatoon, Saskatchewan S7N 5E2, Canada}

\begin{abstract}
The effective potential for radiatively broken electroweak symmetry in the
single Higgs doublet Standard Model is explored to four sequentially
subleading logarithm-summation levels (5-loops) in the dominant Higgs
self-interaction couplant $\lambda $. We augment these results with all
contributing leading logarithms in the remaining large but sub-dominant
Standard Model couplants ($t$-quark, QCD and $SU(2)\otimes U(1)$ gauge
couplants) as well as next to leading logarithm contributions from the
largest of these, the $t$-quark and QCD couplants. Order-by-order stability
is demonstrated for earlier leading logarithm predictions of an $\mathcal{O}$%
($220$ GeV) Higgs boson mass in conjunction with fivefold enhancement of the
value for $\lambda $ over that anticipated from conventional spontaneous
symmetry breaking.
\end{abstract}

\maketitle

Radiative symmetry breaking, as proposed by S.\ Coleman and E.\ Weinberg %
\cite{3}, embraced the premise that the Standard Model (SM) Lagrangian was
protected by some symmetry from a tree level mass term for the Higgs field.
In the absence of large destabilizing Yukawa couplings (heavy fermions),
Coleman and Weinberg were able to show that $\lambda $, the scalar
self-coupling within that Lagrangian, was of the same order as the $%
SU(2)\otimes U(1)$ gauge coupling constants to the fourth power. However,
such a small magnitude for $\lambda $ no longer occurs in the presence of
the large $t$-quark Yukawa coupling, which requires an even larger value of $%
\lambda $ to stabilize the SM effective potential. Very recent work, \cite%
{1,2} based upon full consideration of all contributing leading-logarithm ($%
LL$) terms in the effective potential for the single-Higgs-doublet SM
effective potential that devolves from a Coleman-Weinberg
(mass-term-protected) tree potential, has predicted a Higgs boson mass (218
GeV) well within indirect-measurement bounds \cite{4}, in conjunction with a
much larger scalar-field self-coupling ($y=\lambda /4\pi ^{2}=0.054$) than
would be expected from conventional spontaneous symmetry breaking
[conventionally, $y=m_{H}^{2}/(8\pi ^{2}<\phi >^{2})=0.01$], an enhancement
directly measurable in processes such as $WW\rightarrow HH$ \cite{5}.

The purpose of the present article is to ascertain whether such clear
phenomenological signatures for radiative electroweak symmetry breaking
based upon such a conformally invariant tree potential (as may arise from a
protective higher symmetry) persist upon inclusion of subsequent-to-leading
logarithm contributions to the effective potential.

In ref.\ \cite{1}, the summation of $LL$ contributions to the SM\ effective
potential is expressed in terms of its dominant three couplants $%
x=g_{t}^{2}(v)/4\pi ^{2}=0.0253,\;y=\lambda (v)/4\pi ^{2},\;z=\alpha
_{s}(v)/\pi =0.0329$, where the momentum scale $v=\langle\phi\rangle=246.2\,{\rm GeV}$ is the
vacuum expectation value (vev) of electroweak symmetry breaking. This $LL$
effective potential may be expressed as a power series in the logarithm $%
L=\log (\phi ^{2}/v^{2}):$ 
\begin{equation}
V_{eff}\equiv\pi^2\phi^4S=\pi ^{2}\phi ^{4}\left( A+BL+CL^{2}+DL^{3}+EL^{4}+\ldots \right) .
\label{eq1}
\end{equation}%
The constant $A=y+K$, where $K$ includes all finite $\phi ^{4}$ counterterms
remaining after divergent contributions from $\phi ^{4}$ graphs degree-2 and
higher in couplant powers are cancelled. Coefficients $\{B,C,D,E\}$ are
explicitly obtained in refs.\ \cite{1,2} via the renormalization group (RG)
equation,%
\begin{equation}
\left[ \left( -2-2\gamma \right) \frac{\partial }{\partial L}+\beta _{x}%
\frac{\partial }{\partial x}+\beta _{y}\frac{\partial }{\partial y}+\beta
_{z}\frac{\partial }{\partial z}-4\gamma \right] S=0,  \label{eq2}
\end{equation}%
as degree $\left\{ 2,3,4,5\right\} $ polynomials in the couplants $x$, $y$, $%
z$. The unknown couplant $y(v)$ and finite counterterm $K\phi ^{4}$ are
numerically determined by the simultaneous application of Coleman and
Weinberg's renormalization conditions \cite{3} 
\begin{equation}
V_{eff}^{\prime }(v)=0\Longrightarrow K=-B/2-y;\quad
V_{eff}^{(4)}(v)=V_{tree}^{(4)}(v)\Longrightarrow y=\frac{11}{3}B+\frac{35}{3%
}C+20D+16E  ~.\label{eq3}
\end{equation}%
Given the $LL$ expressions for $\{B,C,D,E\}$ in Eqs.\ (8)--(11) of ref.\ %
\cite{1}, one finds that $y=0.05383$, $K=-0.05794$, in which case
coefficients $\{A,B,C,D,E\}$ in the potential (\ref{eq1}) are numerically
determined. The ``running Higgs boson mass'' [6] at the vev-momentum-scale
is found from the second derivative of the effective potential 
\begin{equation}
m_{H}^{2}=V_{eff}^{\prime \prime }(v)=8\pi ^{2}v^{2}(B+C),  \label{eq4}
\end{equation}%
to be $m_{H}=216$ GeV at $LL$ level \cite{1,2}. Incorporation of $LL$
contributions to $\left\{ A-E\right\} $ from the much smaller electroweak
gauge couplants $r=g_{2}^{2}/4\pi ^{2}=0.0109$, $s=g^{\prime 2}/4\pi
^{2}=0.00324$ modifies the value of $m_{H}$ to be $218$ GeV and $\ y(v)$ to
be $0.0545$ \cite{2}.

We consider here whether this large value of the self-interaction couplant $%
\ y(v)$\ is still sufficiently small for the $LL$\ Higgs boson mass ($218$\
GeV)\ to be subject to controllable corrections from those
subsequent-to-leading-logarithm contributions to the effective potential
that are dominated by higher powers of $y$. To address stability when $y(v)$%
\ is large, we first consider the scalar field theory projection (SFTP) of
the SM effective potential, obtained by setting all SM couplants except the
dominant couplant $y$\ ($y>z,x,r,s)$\ to zero. Indeed, focusing on the
large-coupling subtheory and then taking into account subdominant couplings
is analogous to the usual treatment of processes in which QCD and
electroweak corrections both occur (e.g. $e^{+}e^{-}\rightarrow $ hadrons).

When supplemented by scalar-field kinetic terms, the SFTP of the SM
electroweak effective potential is equivalent to a globally 
${{O}}(4)$ symmetric massless scalar field theory for which $\beta _{y}$
and $\gamma $ have been calculated in $\overline{\mathrm{MS}}$ \cite{8} to 
\textit{five-loop order}. [The coupling constant in ref.\ \cite{8} is $%
g=3\lambda /(8\pi ^{2})=3y/2$.] The $LL$ SFTP of the SM effective potential
is just the $x=0$ limit of Eq.\ (6.1) of ref.\ \cite{2}: 
\begin{equation}
V_{SFTP}^{LL}=\pi ^{2}\phi ^{4}\left[ \frac{y}{1-3yL}+K^{\prime }\right] .
\label{eq5}
\end{equation}%
The constant $K^{\prime }$ represents the contribution of all finite $\phi
^{4}$ counterterms degree-2 and higher in $y$. Curiously, if $v$ retains its
246 GeV SM value, the running Higgs mass and scalar couplant $y(v)$
extracted from this potential are not very different from those of the full $%
LL$ series. By applying conditions (\ref{eq3}) to the SFTP $LL$ values $%
\{B,C,D,E\}=\{3y^{2},9y^{3},27y^{4},81y^{5}\}$ obtained from Eq.\ (\ref{eq5}%
), one finds that $y(v)=0.05414$ and $K^{\prime }=-y-\frac{3}{2}%
y^{2}=-0.05853$. Substituting this value for $y$ into $B$ and $C$\ [Eq.\ (%
\ref{eq4})], we obtain a running Higgs boson mass of 221 GeV at the
vev-momentum-scale, only a small departure from the 216 GeV result \cite{1}
when Standard Model couplants $x$ and $z$ are assigned physical
vev-momentum-scale values instead of the value zero. [Note that the SFTP is 
\textit{not} scale-free; a physical vev scale $v=(\sqrt{2}G_{F})^{-1/2}$
arises from the SM gauge sector.] These results suggest that the SFTP of the
SM \textit{dominates} radiative electroweak symmetry breaking, subject to
manageably small corrections from the other smaller SM interaction couplants
($x,z$, etc.)

In the absence of an explicit mass term, the SFTP \textit{all-orders}
potential takes the form of a perturbative field theoretic series ($%
y=\lambda /4\pi ^{2}$, ${\mathcal{L}}=\log (\phi ^{2}/\mu ^{2})$) 
\begin{equation}
V_{SFTP}=\pi ^{2}\phi ^{4}S_{SFTP},\quad S_{SFTP}=y+\sum_{n=1}^{\infty
}\sum_{m=0}^{n}\;T_{n,m}\;y^{n+1}{\mathcal{L}}^{m}.  \label{eq6}
\end{equation}%
$LL$ contributions to this series involve coefficients $T_{n,n}=3^{n}$ from
Eq.\ (\ref{eq5}); $NLL$ contributions correspond to coefficients $T_{n,n-1}$%
; and so forth. The invariance of $V_{SFTP}$ under changes in the
renormalization scale $\mu $ implies that $S_{SFTP}$ satisfies the ${%
\overline{\mathrm{MS}}}$ renormalization-group (RG) equation (\ref{eq2})
with $\beta _{x}=\beta _{z}=0$, and with \cite{8} 
\begin{equation}
\beta _{y}=6y^{2}-\frac{39}{2}y^{3}+\frac{10332}{55}%
y^{4}-2698.27y^{5}+47974.7y^{6}+\ldots ~,\quad \gamma =\frac{3}{8}y^{2}-%
\frac{9}{16}y^{3}+\frac{585}{128}y^{4}-49.8345y^{5}+\ldots  \label{eq7}
\end{equation}%
The series $S_{SFTP}$ in the full potential (\ref{eq6}) may be rewritten in
terms of \textit{sums} of leading ($S_{0}$) and successively subleading ($%
S_{1},S_{2},...$) logarithms: 
\begin{equation}
S_{SFTP}=yS_{0}(y{\mathcal{L}})+y^{2}S_{1}(y{\mathcal{L}})+y^{3}S_{2}(y{%
\mathcal{L}})+y^{4}S_{3}(y{\mathcal{L}})+\ldots ;\quad S_{k}(u)\equiv
\sum_{n=k}^{\infty }T_{n,n-k}u^{n-k}.  \label{eq9}
\end{equation}

Given $u=y{\mathcal{L}}$, we employ the methods of ref.\ \cite{9} to obtain
successive differential equations for $S_{k}\left( u\right) ,$ first by
substituting Eq. (\ref{eq9}) into the RG equation (\ref{eq2}) with nonzero
RG functions (\ref{eq7}), and then by organizing the RG equation in powers
of $y$:%
\begin{gather}
{\mathcal{O}}\left( y^{2}\right) :\quad 2\left( 1-3u\right) \frac{dS_{0}}{du}%
-6S_{0}=0,\;S_{0}\left( 0\right) =1;  \label{eq10} \\
{\mathcal{O}}\left( y^{3}\right) :\quad 2\left( 1-3u\right) \frac{dS_{1}}{du}%
-12S_{1}=-21S_{0}-\frac{39}{2}u\frac{dS_{0}}{du},\quad S_{1}\left( 0\right)
=T_{1,0};  \label{eq11} \\
{\mathcal{O}}\left( y^{4}\right) :\quad 2\left( 1-3u\right) \frac{dS_{2}}{du}%
-18S_{2}=\frac{41823}{220}S_{0}-\frac{3}{4}\frac{dS_{0}}{du}+\frac{10332}{55}%
u\frac{dS_{0}}{du}-\frac{81}{2}S_{1}-\frac{39}{2}u\frac{dS_{1}}{du},\quad
S_{2}\left( 0\right) =T_{2,0} ~. \label{eq13}
\end{gather}%
Equations for $S_{3}$ and $S_{4}$ (not displayed) are also straightforward
to obtain from the 5-loop RG functions (\ref{eq7}). One can thus obtain
exact solutions to the sums of $LL$, $NLL$, $N^{2}LL$, $N^{3}LL$, and $%
N^{4}LL$ contributions to the series (\ref{eq9}).

As before we choose $\mu =v$, the scalar field $vev$, in which case ${%
\mathcal{L}}\rightarrow L=\log\left( \phi ^{2}/v^{2}\right)$ \cite{10}. Recall that the
counterterm $K^{\prime }$ in the $LL$ potential (\ref{eq5}) ($K^{\prime
}=-0.05853$) and the couplant $y\;(=0.05414)$ are comparable in magnitude; $%
K^{\prime }$ is \textit{not} a single higher-order counterterm coefficient.
The SFTP $\phi ^{4}$ counterterm coefficient $K^{\prime }$ can accommodate
contributions of any logarithm-free terms $T_{n,0}y^{n+1}\;\;(n>1)$ in the
complete series (\ref{eq6}). The all-orders SFTP of the effective potential 
\begin{equation}
V_{SFTP}=\pi ^{2}\phi ^{4}y\sum_{n=0}^{\infty
}\sum_{m=0}^{n}T_{n,m}y^{n}L^{m}=\pi ^{2}\phi ^{4}y\sum_{k=0}^{\infty
}y^{k}S_{k}(yL),  \label{eq15}
\end{equation}%
is then approached by the following successive approximations to Eq.\ (\ref%
{eq15}), which incorporate the summations of successively subleading
logarithms contributing to the complete series (\ref{eq9}): 
\begin{gather}
V_{LL}=\pi ^{2}\phi ^{4}y\left[ \sum_{n=0}^{\infty
}T_{n,n}(yL)^{n}+yT_{1,0}+y^{2}T_{2,0}+y^{3}T_{3,0}+\ldots \right] \equiv
\pi ^{2}\phi ^{4}[yS_{0}(yL)+K^{\prime }],  \label{eq16} 
\\
\begin{split}
V_{NLL}&=\pi ^{2}\phi ^{4}y\left[ \sum_{n=0}^{\infty
}T_{n,n}(yL)^{n}+y\sum_{n=1}^{\infty }T_{n,n-1}(yL)^{n-1}+y^{2}T_{2,0}+\ldots %
\right]  \\
&=\pi ^{2}\phi ^{4}\left[ yS_{0}(yL)+y^{2}S_{1}(yL)+(K^{\prime }-y^{2}T_{1,0})%
\right]  ~,
\end{split}
\label{eq17}  
\\
V_{N^{p}LL}=\pi ^{2}\phi ^{4}\left[ y\sum_{q=0}^{p}y^{q}S_{q}(yL)+\left(
K^{\prime }-\sum_{q=1}^{p}y^{q+1}T_{q,0}\right) \right] ,\overset{lim}{%
_{p\rightarrow \infty }}V_{N^{p}LL}=V_{SFTP}.  \label{eq18}
\end{gather}%
Note from Eq.\ (\ref{eq16}) that $K^{\prime }$\ is numerically inclusive of 
\textit{all} finite $\phi ^{4}$ counterterms. Thus for the $NLL$ case with $%
K^{\prime }$ already determined from application of Eq. (\ref{eq3}) to Eq. (%
\ref{eq5}), we now find from Eq.\ (\ref{eq3}) that $T_{1,0}=-\left[
4K^{\prime }-21y^{3}+6y^{2}+4y\right] /12y^{3}=2.5521$, $y=0.05381$, and
from Eq.\ (\ref{eq4}) that $\left[ V_{SFTP}^{\prime \prime }(v)\right]
^{1/2}=227$ GeV.

One can continue this procedure through $N^{4}LL$ order, applying conditions
(\ref{eq3}) on potentials (\ref{eq18}) to determine $T_{n,0}$ and $y$ while
making use of the information ($T_{1,0},\ldots ,T_{n-1,0};K^{\prime }$) from
preceding orders. The results, as summarized in columns 2-4 of Table \ref%
{tab1}, show remarkable order-by-order stability in the values obtained both
for the couplant $y(v)$ and the running Higgs boson mass. Note also that the
SFTP potential (\ref{eq15}) is compatible \emph{by construction} with the $%
\overline{\mathrm{MS}}$ renormalization scheme, since the summations $%
S_{k}(yL)$ of $\mathrm{N^{k}LL}$'s are obtained from differential equations [%
\textit{e.g.\ }Eqs.\ (\ref{eq10})--(\ref{eq13})] derived from $\overline{%
\mathrm{MS}}$ RG functions (\ref{eq7}).

\begin{table}[tbp]
\centering
\begin{tabular}{||c|c|c|c||c|c|c||c|c|c||}
\hline
\multicolumn{4}{||c||}{$N^{n}LL\;\;$SFTP} & \multicolumn{3}{|c||}{SFTP $+LL$\
in $\left\{ x,z\right\} $} & \multicolumn{3}{|c||}{SFTP $+LL$\
in $\left\{ x,z,r,s\right\} $}\\ \hline
$n$ & $y\left( v\right) $ & $m_{H}$ & $T_{n,0}$ & $y\left( v\right)$ & $
m_{H}$ & $T_{n,0}$ &  $y\left( v\right)$ & $
m_{H}$ & $T_{n,0}$\\ \hline
$0$ & $0.05414$ & $221.2$ & $1$ & $0.05383$ & $215.8$ & $1$ & $0.05448$ & $218.3$ & $1$  
\\ \hline
$1$ & $0.05381$ & $227.0$ & $2.5521$ & $0.05351$ & $221.7$ & $2.5533$ & $0.05415$ & $224.4$ & $2.5603$
\\ \hline
$2$ & $0.05392$ & $224.8$ & $-8.1770$ & $0.05362$ & $219.5$ & $-8.1744$ & $0.05426$ & $222.1$ & $-8.1773$ 
\\ \hline
$3$ & $0.05385$ & $226.2$ & $83.211$ & $0.05355$ & $221.3$ & $83.190$ & $0.05419$ & $224.0$ & $83.195$ \\ \hline
$4$ & $0.05391$ & $225.0$ & $-1141.8$ & $0.05338$ & $223.6$ & $-982.21$ & $0.05406$ & $225.5$ & $-1191.8$ \\ \hline
\end{tabular}%
\caption{ Perturbative stability of results inclusive of $N^{n}LL$
contributions from the dominant couplant $y\left( v\right) =\left( \protect%
\lambda \left( v\right) /4\protect\pi ^{2}\right) $ to the SFTP of the SM
effective potential (columns 2-4). Columns 5--7 show the effect of
augmenting this projection with prior determinations of $LL$ contributions
to the effective potential from $t$-quark $\left( x=g_{t}^{2}\left( v\right)
/4\protect\pi ^{2}\right) $ and from QCD $\left( z=\protect\alpha _{s}\left(
v\right) /\pi\right) $. Columns 8--10 further augment this projection with $LL$
contributions from electroweak $SU\left( 2\right) $ $%
\left( r\equiv g_{2}^{2}\left( v\right) /4\protect\pi ^{2}\right) $ and $%
U\left( 1\right) \left( s\equiv g^{\prime 2}/4\protect\pi ^{2}\right) $
gauge couplants. $m_{H}$ denotes the vev-referenced running Higgs boson mass 
$\left[ V_{eff}^{\prime \prime }\left( v\right) \right] ^{1/2}$ in GeV units 
$\left( v=246.2\;GeV\right) $. }
\label{tab1}
\end{table}

We now augment the SFTP with the smaller subdominant SM\ couplants $\left\{
x,z,r,s\right\} $ \cite{1,2}. 
If only the $LL$ from $\{x,z\}$, the Yukawa interaction sector, are included, $K$ is found from the minimization condition (\ref{eq3}) to be $K=-y-3y^2/2+3x^2/8=-0.05793$, where $y$ is the $LL$ value $0.05383$ obtained [via Eq.\ (\ref{eq3})] in refs.\ \cite{1,2}. If $LL$ contributions from $\{r,s\}$, the $SU(2)\otimes U(1)$ gauge couplants, are also included, the  
constant $K$ is now 
$K=-y-3y^{2}/2+3x^{2}/8-3rs/64-9r^{2}/128$ $-3s^{2}/128=-0.058704$, where $y$
is the $LL$ value for $y\left( v\right) =0.054481$ obtained [via Eq.\ (\ref%
{eq3})] in ref.\ \cite{2}. These results are listed in columns 5--10 of
Table \ref{tab1}. The order-by-order stability of improved predictions $y\left(
v\right) =0.054$, $m_{H}=220\mbox{--}227\,\mathrm{GeV}$ is quite striking. Moreover, we
have also found that this stability persists when contributions from
subdominant couplants $x$ and $z$ are considered to $NLL$ order via use of
two-loop $\overline{\mathrm{MS}}$ RG functions \cite{11} in Eq.\ (\ref{eq2}%
). The all-orders effective potential analogous to Eq.\ (\ref{eq6}) but now
inclusive of all three dominant SM couplants $x,~y,~z$ is of the form 
\begin{equation}
V_{xyz}=\pi ^{2}\phi ^{4}\sum_{n=0}^{\infty }x^{n}\sum_{k=0}^{\infty
}y^{k}\sum_{\ell =0}^{\infty }z^{\ell }\sum_{p=0}^{n+k+\ell
-1}L^{p}D_{n,k,\ell ,p}=\pi ^{2}\phi ^{4}S_{xyz}\quad \left(
D_{0,1,0,0}=1,~D_{1,0,0,0}=D_{0,0,1,0}=0\right) ,  \label{v23}
\end{equation}%
where the series $S_{xyz}$ can be expressed either as summations of $LL$s ($%
p=n+k+l-1$), $NLL$s ($p=n+k+l-2$),\textit{etc}, as in Eq.\ (\ref{eq15}), or
as power series in the logarithm $L$, as in Eq.\ (\ref{eq1}). To $NLL$
order, only the $y^{2}$ ($D_{0,2,0,0}=T_{1,0}\equiv a$) and $x^{2}$ ($%
D_{2,0,0,0}\equiv b$) finite counterterms from divergent one-loop $\phi ^{4}$
graphs contribute to the coefficients $\{B,C,D,E\}$. The $NLL$ contributions
to $\{B,C,D,E\}$ from Eq.\ (\ref{v23}) are {\allowdisplaybreaks 
\begin{gather}
\begin{split}
B=&\left[3y^2 -\frac{3}{4}x^{2}\right]_{LL}\\
&+\left[ \left( -\frac{27}{4}+\frac{3a%
}{2}\right) xy^{2}+\left( \frac{3}{2}+\frac{3b}{4}\right) x^{3}-\left(
1+4b\right) x^{2}z+\left( 6a-\frac{21}{2}\right) y^{3}+\left( \frac{3}{4}-%
\frac{3a}{2}\right) x^{2}y\right] _{NLL},  
\end{split}
\label{bnll} \\
\begin{split}
C=& \left[ 9y^{3}+\frac{9}{4}xy^{2}-\frac{9}{4}x^{2}y+\frac{3}{2}x^{2}z-%
\frac{9}{32}x^{3}\right] _{LL} \\
& +\biggl[\left( 27a-\frac{621}{8}\right) y^{4}+\left( \frac{27a}{2}-\frac{%
225}{4}\right) xy^{3}+\left( -\frac{3a}{2}+\frac{21}{2}\right)
xy^{2}z+\left( 3a-\frac{9}{2}\right) x^{2}yz+\left( -\frac{225a}{32}+\frac{27%
}{8}\right) x^{2}y^{2}\biggr. \\
\biggl.& \qquad +\left( \frac{23b}{2}+\frac{127}{16}\right)
x^{2}z^{2}+\left( -\frac{27}{4}-\frac{15b}{4}\right) x^{3}z+\left( -\frac{45a%
}{16}+\frac{351}{32}\right) x^{3}y+\left( \frac{405}{256}+\frac{45b}{64}+%
\frac{9a}{16}\right) x^{4}\biggr]_{NLL},
\end{split}
\label{cnll} \\
\begin{split}
D =& \left[ 27y^4 + \frac{27}{2} xy^3 - \frac{3}{2} xy^2 z + 3x^2yz - \frac{%
225}{32} x^2 y^2 - \frac{23}{8} x^2 z^2 + \frac{15}{16} x^3 z - \frac{45}{16}
x^3 y + \frac{99}{256} x^4\right]_{LL} \\
&+\Biggl[ \left( -\frac{801}{2} + 108 a \right) y^5 + \left( -\frac{11547}{32%
} + 81 a \right) xy^4 + \left( -12 a + \frac{147}{2} \right) xy^3 z\Biggr. +
\left( \frac{15 a}{8} - \frac{291}{16} \right) xy^2 z^2 \\
&\qquad+ \left( - \frac{23 a}{4}+ \frac{75}{4}\right) x^2 yz^2 + \left( -%
\frac{45}{32} - \frac{45 a}{2} \right) x^2 y^3 + \left( \frac{177 a}{16} - 
\frac{33}{8} \right) x^2 y^2 z + \left( - \frac{877}{32} - \frac{115 b}{4}
\right) x^2 z^3 \\
&\qquad+ \left( \frac{3125}{128} + \frac{201 b}{16} \right) x^3 z^2 + \left( 
\frac{69 a}{8} - \frac{615}{16} \right) x^3 yz + \left( \frac{19323}{256} - 
\frac{2781 a}{128} \right) x^3 y^2 \\
&\qquad + \left( -\frac{1023}{128} - \frac{135 b}{32} - \frac{9 a}{4}
\right) x^4 z + \Biggl. \left( \frac{3825}{256} + \frac{45 a}{128} \right)
x^4 y + \left( -\frac{1035}{512} + \frac{45 b}{64} + \frac{81 a}{64} \right)
x^5\Biggr]_{NLL},
\end{split}
\label{dnll} \\
\begin{split}
E =& \Biggl[81y^5 + \frac{243}{4} xy^4 - 9xy^3 z + \frac{45}{32} xy^2 z^2 - 
\frac{69}{16} x^2 yz^2 - \frac{135}{8} x^2 y^3 + \frac{531}{64} x^2 y^2 z + 
\frac{345}{64} x^2 z^3 - \frac{603}{256} x^3 z^2 \Biggr. \\
& + \Biggl.\frac{207}{32} x^3 yz - \frac{8343}{512} x^3 y^2 - \frac{459}{512}
x^4 z + \frac{135}{512} x^4 y + \frac{837}{1024} x^5\Biggr]_{LL} \\
&+\Biggl[\left( - \frac{55539 a}{4096} + \frac{1081377}{8192} \right) y^2
x^4 + \left( \frac{2187 a}{256} - \frac{29133}{2048} \right)yx^5 + \left( - 
\frac{125793}{64} + 405 a \right) y^5 x \\
& \qquad+ \left( \frac{1035 b}{64} + \frac{105 a}{16} + \frac{111633}{4096}
\right) x^4 z^2 + \left( - \frac{1215 a}{32} - \frac{195939}{1024} \right)
y^4 x^2 + \left( \frac{4255 b}{64} + \frac{38613}{512} \right) x^2 z^4 \\
& \qquad + \left( -\frac{315 b}{64} - \frac{207 a}{32} + \frac{17703}{2048}
\right) x^5 z + \left( -\frac{4509 b}{128} - \frac{75315}{1024} \right) x^3
z^3 + \left( - \frac{28323}{16} + 405 a \right) y^6 \\
&\qquad + \left( \frac{4581}{64} + \frac{855 a}{32}\right) y^3 x^2 z +
\left( - \frac{31455 a}{256} + \frac{227529}{512} \right) y^3 x^3 + \left( 
\frac{3231}{8} - \frac{135 a}{2} \right) y^4 xz \\
& \qquad+ \left( -\frac{3807}{32} + \frac{225 a}{16}\right) y^3 xz^2 +
\left( -\frac{28197}{128} + \frac{7191 a}{128} \right) y^2 x^3 z + \left( 
\frac{14847}{512} - \frac{1215 a}{64} \right) y^2 x^2 z^2 \\
&\qquad + \left( \frac{3603}{128} - \frac{165 a}{64}\right) y^2 xz^3 +
\left( - \frac{35145}{512} + \frac{621 a}{256}\right) yx^4 z + \left( \frac{%
7323}{64}- \frac{2643 a}{128}\right) yx^3 z^2 \\
& \qquad+ \Biggl. \left( -\frac{93}{2} + \frac{345 a}{32} \right) yx^2 z^3 +
\left( - \frac{208629}{32768} + \frac{1485 b}{2048} + \frac{1269 a}{1024}
\right) x^6\Biggr]_{NLL}.
\end{split}
\label{enll}
\end{gather}%
} 
Analogous to $A=y+K^{\prime }$\ in the series expansions of Eqs.\ (\ref{eq16}%
)--(\ref{eq18}), the leading term $A$ in the power series (\ref{eq1}) is
just $y+K$, where the constant $K$ is inclusive of all degree-2 and higher
purely $\phi ^{4}$ terms ($p=0$) in the full potential (\ref{v23}). Given
our previous determination (in the absence of gauge couplants $r$ and $s$)
of $K=-0.057935 $ \cite{1,2} and our $NLL$ ${\rm SFTP}+\{x,z\}_{LL}$ result $a=T_{1,0}=2.5533$ [Table \ref{tab1}] we find
upon application of conditions (\ref{eq3}) that $b=-17.306$ and $y(v)=0.05311$%
. Substituting these results into Eq.\ (\ref{eq4}) [via Eqs.\ (\ref{bnll})--(%
\ref{enll})], we find that $V_{eff}^{\prime \prime }=(227.8\,\mathrm{GeV}%
)^{2}$. 
This result, involving a full $NLL$ treatment of dominant $\{x,y,z\}$ contributions to $V_{eff}$, is a full next-order extension  of the $LL$ contributions of $\{x,y,z,\}$ to $V_{eff}$ presented in \cite{1}.
If we further augment this $NLL$ result with $LL$ contributions from
electroweak gauge couplants \cite{2}, 
$T_{1,0}=2.5603$ [Table \ref{tab1}], $b=-17.857$,
 $y(v)=0.05374$, and $%
V_{eff}^{\prime \prime }=(230.7\,\mathrm{GeV})^{2}$.

In radiative electroweak symmetry breaking, the next order relationship between the
physical Higgs boson mass and $V^{\prime\prime}_{eff}$ has been worked out
in principle \cite{ms}. In particular, the next order Higgs inverse propagator mass
term must remain $V^{\prime\prime}_{eff}(v)$ because of the absence of a
primitive $\phi^2$ term in the original Lagrangian. The kinetic term for the
inverse propagator (as seminally discussed for the massless gauge boson propagator 
in \cite{pol}) must retain consistency with the
relation $\mu d\phi/d\mu=-\gamma(\mu)\phi$ implicit within the RG equation (%
\ref{eq2}), in which case the next-order inverse propagator for the Higgs
field at $\mu=v$ may be expressed as 
\begin{equation}
\Gamma\left(p^2,v\right)=\left[1-\left(\frac{3}{4}x(v)-\frac{9}{16}r(v)-
\frac{3}{16}s(v)\right) \log{\left(\frac{p^2}{v^2}\right)} \right]p^2
-V^{\prime\prime}_{eff}(v)~.  \label{inv}
\end{equation}
We have included only those SM contributions to $\gamma(v)$ that that are
linear in the couplants $\{x,y,z,r,s\}$ \cite{11}. The zero of (\ref{inv})
is the NLL prediction for the physical Higgs boson mass [$%
\Gamma\left(m_H^2,v\right)=0$], which is found to be reduced by only 
$0.2\mbox{--}0.3\,\mathrm{GeV}$ from 
values respectively below $231\,{\rm GeV}$ and above $220\,{\rm GeV}$ extracted from
$V^{\prime\prime}_{eff}(v)$  past $LL$ order.
We therefore conclude that the
$220\mbox{--}230\,\mathrm{GeV}$  Higgs boson mass and a factor of five
enhancement of the scalar-field self-interaction coupling are indeed
signature predictions for \textit{radiative} SM electroweak symmetry
breaking. This enhancement should be particularly evident in $WW\to HH$ cross-sections \cite{5} accessible in the not
too distant future.

\begin{acknowledgments}
We are grateful for useful discussions with F.A.\ Chishtie, M.\ Sher and
V.A.\ Miransky and for support from the Natural Sciences and Engineering
Research Council of Canada.
\end{acknowledgments}


\begin{thebibliography}{99}
\bibitem{10} The approximate scale invariance of $LL$ results is discussed
in refs.\ [2,3].
\end{thebibliography}
\end{document}